\documentclass[journal=jctcce,manuscript=article]{achemso}
\usepackage{commath}
\usepackage[section]{placeins}

\SectionNumbersOn


\title{Particle-Particle Random Phase Approximation for Predicting Correlated Excited States of Point Defects}

\author{Jiachen Li}
\affiliation{Department of Chemistry, Yale University, New Haven, Connecticut 06520, United States}
\email{jiachen.li@yale.edu}
\author{Yu Jin}
\affiliation{Pritzker School of Molecular Engineering, University of Chicago, Chicago, Illinois 60637, United States}
\author{Jincheng Yu}
\affiliation{Department of Chemistry, Duke University, Durham, North Carolina 27708, United States}
\author{Weitao Yang}
\affiliation{Department of Chemistry, Duke University, Durham, North Carolina 27708, United States}
\email{weitao.yang@duke.edu}
\author{Tianyu Zhu}
\affiliation{Department of Chemistry, Yale University, New Haven, Connecticut 06520, United States}
\email{tianyu.zhu@yale.edu}

\begin{document}

\begin{tocentry}
\includegraphics[width=1\textwidth]{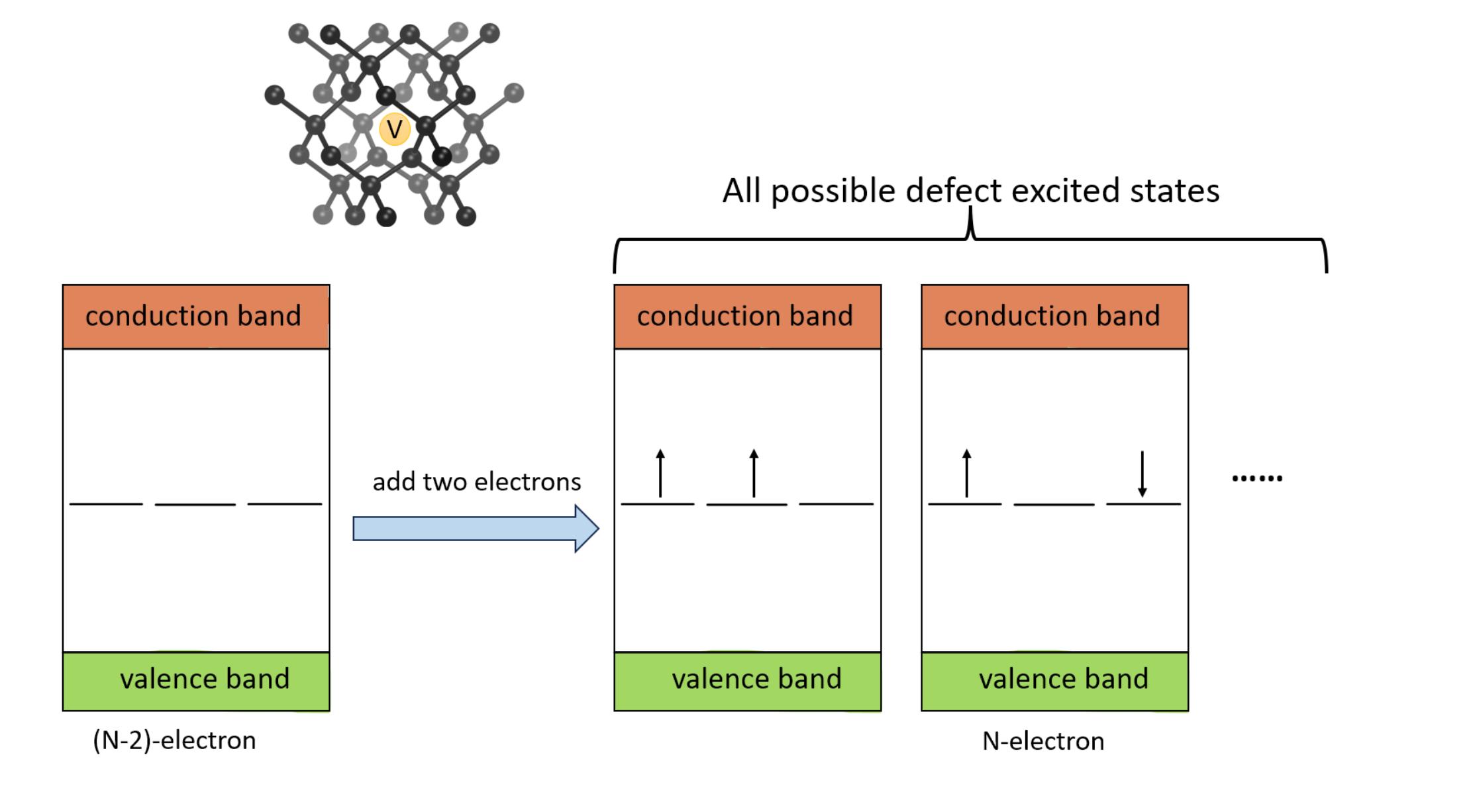}
\end{tocentry}

\begin{abstract}
The particle-particle random phase approximation (ppRPA) within the hole-hole channel was recently proposed as an efficient tool for computing excitation energies of point defects in solids [\textit{J. Phys. Chem. Lett.} 2024, 15, 2757-2764]. In this work, we investigate the application of ppRPA within the particle-particle channel for predicting correlated excited states of point defects, including the carbon-vacancy (VC) in diamond, the oxygen-vacancy (VO) in magnesium oxide (MgO), and the carbon dimer defect (C$_{\text{B}}$C$_{\text{N}}$) in two-dimensional hexagonal boron nitride (h-BN). Starting from a density functional theory calculation of the ($N-2$)-electron ground state, vertical excitation energies of the $N$-electron system are obtained as the differences between the two-electron addition energies. We show that active-space ppRPA with the B3LYP functional yields accurate excitation energies, with errors mostly smaller than 0.1 eV for tested systems compared to available experimental values. We further develop a natural transition orbital scheme within ppRPA, which provides insights into the multireference character of defect states. This study, together with our previous work, establishes ppRPA as a low-cost and accurate method for investigating excited-state properties of point defect systems.
\end{abstract}

\maketitle

\section{INTRODUCTION}

Optically active point defects in semiconductors and insulators are promising platforms for quantum information processing due to their special optical and magnetic properties\cite{dreyerFirstPrinciplesCalculationsPoint2018,wolfowiczQuantumGuidelinesSolidstate2021}.
In many point defect systems,
defect energy levels are introduced within the fundamental gap of the host material,
and the transitions among them can absorb or emit photons at lower energies than the optical gap of the pristine host material.
%then an incident photon can be excited by the light absorbed at lower energies.
The properties of the pristine host material are thus significantly changed, 
offering distinct advantages in applications of quantum information science\cite{jelezkoObservationCoherentOscillation2004,weberQuantumComputingDefects2010,andersonFivesecondCoherenceSingle2022}.
These point defects may function as quantum bits (qubits), 
%which enables an optical spin-polarization of the electron spin state to be initialized and readout via spin-selective non-radiative decay from its optical excited state
and the qubit initialization and readout are enabled by an optical spin-polarization cycle involving both radiative and spin-selective non-radiative transitions between many-body ground and excited states \cite{childressCoherentDynamicsCoupled2006,galiInitioTheoryNitrogenvacancy2019}.
To assist in the interpretation of experiment results and gain deeper insights into materials design,
theoretical simulations play an important role in the identification and characterization of point defects.

In past decades,
many efforts have been devoted to developing theoretical approaches for the accurate description of excited states of point defect systems.
Because of the good balance between the computational cost and accuracy, 
density functional theory (DFT)~\cite{hohenbergInhomogeneousElectronGas1964,kohnSelfConsistentEquationsIncluding1965,parrDensityFunctionalTheoryAtoms1989} based methods have been widely used to study point defects.
Linear-response time-dependent density functional theory (TDDFT)~\cite{casidaTimeDependentDensityFunctional1995,ullrichTimeDependentDensityFunctionalTheory2011} and $\Delta$SCF~\cite{zwijnenburgOpticalExcitationsDefects2008,mackoit-sinkevicieneCarbonDimerDefect2019,jinPhotoluminescenceSpectraPoint2021,mackrodtCalculatedEnergiesCharge2022} have been the most popular tools for calculating excited-state properties of point defects,
such as excitation energies, geometry relaxations of excited states, and optical absorption and emission spectra~\cite{jinExcitedStateProperties2023,jinVibrationallyResolvedOptical2022}. 
However,
it is challenging for methods based on a single Kohn-Sham (KS) determinant to describe defect states with strong multireference characters\cite{cohenChallengesDensityFunctional2012,zhuManyelectronExpansionDensity2016,zhuImplementationManyPairExpansion2019}. 
To obtain more reliable treatments of the correlated excited states, 
quantum many-body methods including the Bethe-Salpeter equation approach within the Green's function formalism (i.e., $GW$-BSE)\cite{salpeterRelativisticEquationBoundState1951,shamManyParticleDerivationEffectiveMass1966,hankeManyParticleEffectsOptical1979}, 
equation-of-motion coupled-cluster theory~\cite{galloPeriodicEquationofmotionCoupledcluster2021},
and quantum Monte Carlo~\cite{simulaCalculationEnergiesMultideterminant2023} have been applied. Nevertheless, their further applications to large point defect systems are limited either by their high computational costs or inherent single-reference nature.
Recently, many flavors of quantum embedding approaches have been developed to describe correlated defect systems with affordable computational costs.
The quantum embedding formalism provides a natural way to focus computation on a chosen active space representing the localized defect states with accurate but computationally demanding high-level theories,
while the remaining environment is treated by efficient low-level theories\cite{sunQuantumEmbeddingTheories2016,cuiEfficientImplementationInitio2020,zhuEfficientFormulationInitio2020,zhuInitioFullCell2021,Zhu2024Kondo,Li2024ibdet}.
It has been shown that the quantum defect embedding theory (QDET)~\cite{maQuantumSimulationsMaterials2020,
maQuantumEmbeddingTheory2021,shengGreenFunctionFormulation2022}, 
the density matrix embedding theory (DMET)~\cite{mitraExcitedStatesCrystalline2021,haldarLocalExcitationsCharged2023,
vermaOpticalPropertiesNeutral2023},
the constrained random phase approximation (cRPA) 
combined with exact diagonalization (ED)~\cite{bockstedteInitioDescriptionHighly2018,
muechlerQuantumEmbeddingMethods2022},
the dynamical downfolding approach\cite{romanovaDynamicalDownfoldingLocalized2023},
and the regional embedding theory~\cite{lauOpticalPropertiesDefects2024} have achieved mixed successes in simulating point defect systems.

Parallel to particle-hole formalisms such as TDDFT and BSE,
the particle-particle random phase approximation (ppRPA)~\cite{vanaggelenExchangecorrelationEnergyPairing2013,vanaggelenExchangecorrelationEnergyPairing2014} offers another path to compute excitation energies from the particle-particle and hole-hole channels.
ppRPA that is originally developed for treating nuclear many-body interactions\cite{ringNuclearManyBodyProblem2004,ripkaQuantumTheoryFinite1986} has been formulated to describe various properties of molecular systems\cite{vanaggelenExchangecorrelationEnergyPairing2013,vanaggelenExchangecorrelationEnergyPairing2014}.
Compared with particle-hole random phase approximation (phRPA) that describes the response of the density matrix to an external perturbation,
ppRPA describes the response of the pairing matrix to a perturbation in the form of a pairing field,
which leads to two-electron addition and removal energies\cite{vanaggelenExchangecorrelationEnergyPairing2013,vanaggelenExchangecorrelationEnergyPairing2014}.
Therefore,
the neutral excitation energy of an $N$-electron system can be obtained from the energy difference between two-electron addition energies of the corresponding ($N-2$)-electron system,
or the energy difference between two-electron removal energies of the corresponding ($N+2$)-electron system.
It has been shown that ppRPA provides good accuracy for modeling excited-state properties of molecular systems such as excitation energies\cite{yangBenchmarkTestsSpin2013,yangDoubleRydbergCharge2013,yangExcitationEnergiesParticleparticle2014,yangSingletTripletEnergy2015,zhangAccurateEfficientCalculation2016,yangChargeTransferExcitations2017,chenMultireferenceDensityFunctional2017,pinterSpinstateEnergeticsIron2018,al-saadonAccurateTreatmentChargeTransfer2018,liMultireferenceDensityFunctional2022,bannwarthHoleHoleTamm2020,yuInitioNonadiabaticMolecular2020,hohensteinPredictionsPreedgeFeatures2021}, 
oscillator strengths\cite{yangDoubleRydbergCharge2013},
conical intersection\cite{yangConicalIntersectionsParticle2016},
and potential energy surfaces\cite{zhangAnalyticGradientsGeometry2014}.
Along with excited-state properties, 
ppRPA can also be used to calculate ground-state properties such as total energies and geometries\cite{vanaggelenExchangecorrelationEnergyPairing2013,yangBenchmarkTestsSpin2013,vanaggelenExchangecorrelationEnergyPairing2014,zhangAnalyticGradientsGeometry2014,chenMultireferenceDensityFunctional2017,liMultireferenceDensityFunctional2022}.
For ground states,
the ppRPA correlation energy is shown to be equivalent to the ladder-coupled-cluster doubles\cite{pengEquivalenceParticleparticleRandom2013,scuseriaParticleparticleQuasiparticleRandom2013}. Excitation energies obtained from ppRPA can be considered as an approximation to double-electron-affinity or
double-ionization-potential equation-of-motion coupled-cluster doubles\cite{yangDoubleRydbergCharge2013,berkelbachCommunicationRandomphaseApproximation2018}.
In the context of DFT,
ppRPA is the first known functional that captures the energy derivative discontinuity in strongly correlated systems and has no delocalization error in single-bond dissociations\cite{vanaggelenExchangecorrelationEnergyPairing2013,vanaggelenExchangecorrelationEnergyPairing2014}.
In the Green's function formalism,
ppRPA eigenvalues and eigenvectors are used to construct the T-matrix self-energy for predicting quasiparticle energies\cite{zhangAccurateQuasiparticleSpectra2017,liRenormalizedSinglesGreen2021,orlandoThreeChannelsManybody2023,liLinearScalingCalculations2023} and the BSE kernel for neutral excitation energies\cite{loosStaticDynamicBethe2022,orlandoChapterElevenExploring2023,moninoConnectionsPerformancesGreen2023}.

Recently, we employed ppRPA to calculate vertical excitation energies (VEEs) of point defects\cite{liAccurateExcitationEnergies2024},
which is the first application of ppRPA to realistic periodic bulk systems.
In Ref.~\citenum{liAccurateExcitationEnergies2024},
VEEs of point defects are computed from the hole-hole channel in ppRPA,
which is the difference between the lowest and a higher two-electron removal energy of the ($N+2$)-electron ground state.
It shows that ppRPA predicts accurate VEEs of the nitrogen-vacancy (NV$^-$) and the silicon-vacancy (SiV$^0$) centers in diamond and the divacancy center (VV$^0$) in 4H silicon carbide with errors smaller than $0.2$ \,{eV} when using the B3LYP functional~\cite{beckeDensityFunctionalThermochemistry1993,leeDevelopmentColleSalvettiCorrelationenergy1988}.
For these point defects,
the corresponding ($N+2$)-electron systems are closed-shell, and all desired defect excited states can be accessed by removing two electrons.
Besides excellent accuracy,
ppRPA is computationally favorable and cheaper than the corresponding ground-state DFT calculation by combining with the recently developed active-space approach\cite{liLinearScalingCalculations2023}.
The ppRPA excitation energies calculated from the hole-hole channel demonstrate rapid convergence, requiring just a few hundred canonical orbitals in the active space,
which is independent of the specific defect system\cite{liAccurateExcitationEnergies2024}.

In this work, 
we further establish the applicability of ppRPA in predicting correlated excited states of point defects by investigating two important extensions. 
First,
the particle-particle channel in ppRPA is adopted to calculate VEEs of point defect systems,
including the carbon-vacancy (VC) in diamond, 
the oxygen-vacancy (VO) in magnesium oxide (MgO), and the carbon dimer (C$_{\text{B}}$C$_{\text{N}}$) defect in two-dimensional hexagonal boron nitride (h-BN).
Compared with previous work that starts with the ($N+2$)-electron ground state\cite{liAccurateExcitationEnergies2024}, 
this work employs the ($N-2$)-electron ground state computed at the DFT level as the reference in ppRPA calculations.
By removing two electrons from the original $N$-electron defect system,
the ($N-2$)-electron ground state becomes closed-shell for systems studied in this paper,
which can be properly described by single-determinant KS-DFT.
All desired excitation energies can then be obtained by taking the differences between two-electron addition energies of the ($N-2$)-electron system.
To reduce the computational cost,
the ppRPA equation is solved with the Davidson algorithm\cite{yangExcitationEnergiesParticleparticle2014} in the canonical active space\cite{liLinearScalingCalculations2023},
then excitation energies at the full-system limit are obtained by an active-space extrapolation scheme.
We demonstrate that ppRPA predicts accurate excitation energies for all tested defect systems. 
Second, we develop a natural transition orbital (NTO) approach within ppRPA to gain further insights into the character of defect excited states.
In analogy to NTOs in particle-hole methods such as TDDFT and BSE\cite{martinNaturalTransitionOrbitals2003,krauseImplementationBetheSalpeter2017,choSimplifiedGWBSE2022},
NTOs in ppRPA are obtained from the singular value decomposition (SVD) of ppRPA eigenvectors.
Similar to other local orbital approaches that provide the understanding of chemical reactivity\cite{glendeningNBONaturalBond2013,zhangUnravellingChemicalInteractions2018,glendeningResonanceNaturalBond2019,yuDescribingChemicalReactivity2022,longguChemicalConceptsMolecular2024,Zhu2018self},
NTOs in ppRPA provide a compact orbital representation for two-electron addition and removal processes.
The weights of NTOs are used to qualitatively measure the multireference character of defect states.

The remainder of this article is organized as follows.
We review the ppRPA formalism and introduce the NTO approach in ppRPA in Section~\ref{sec:theory}. 
The computational details about our calculations are given in Section~\ref{sec:comp_detail}. 
The active-space extrapolation scheme, 
predicted VEEs of tested point defects, 
and the NTO analysis for the multireference character of defect states are presented in Section~\ref{sec:results}.
We finally draw conclusions in Section~\ref{sec:conclusion}.

\section{THEORY}\label{sec:theory}

\subsection{Excitation Energy from ppRPA}
We first review the ppRPA formalism.
ppRPA can be derived from different approaches, 
including the equation of motion\cite{ringNuclearManyBodyProblem2004,
roweEquationsofMotionMethodExtended1968}, 
the adiabatic connection\cite{vanaggelenExchangecorrelationEnergyPairing2013,
vanaggelenExchangecorrelationEnergyPairing2014}, 
and TDDFT with the pairing field\cite{pengLinearresponseTimedependentDensityfunctional2014}.
As the counterpart of phRPA,
ppRPA describes the instantaneous fluctuation of the pairing matrix\cite{vanaggelenExchangecorrelationEnergyPairing2013,vanaggelenExchangecorrelationEnergyPairing2014}
\begin{equation}
    \kappa (x_1, x_2) = \langle \Psi^N_0 | \hat{\psi} (x_2) \hat{\psi} (x_1) | \Psi^N_0 \rangle
\end{equation}
which is zero for a system with a fixed number of electrons.
Here, $x=(r,\sigma)$ is the space-spin combined variable, 
$\Psi^N_0$ is the $N$-electron ground state, 
$\hat{\psi}^{\dagger}$ and $\hat{\psi}$ are creation and annihilation operators in the second-quantization notation.
In the linear-response theory, 
the paring matrix $\delta \kappa (x_1, x_2)$ is non-zero when the system is perturbed by an external field in the form of a pairing field\cite{vanaggelenExchangecorrelationEnergyPairing2013}.
In the frequency domain, 
the time-ordered pairing matrix fluctuation that describes the linear response of the
pairing matrix is\cite{
vanaggelenExchangecorrelationEnergyPairing2013,
vanaggelenExchangecorrelationEnergyPairing2014} 
\begin{equation}\label{eq:pairing_matrix}
    K_{pqrs}(\omega) = 
        \sum_m \frac{
            \langle \Psi^N_0 | \hat{a}_p \hat{a}_q | \Psi^{N+2}_m \rangle 
            \langle \Psi^{N+2}_m | \hat{a}_s^{\dagger} \hat{a}_r^{\dagger} | \Psi^N_0 \rangle 
            }{
            \omega - \Omega^{N+2}_m + i\eta
            }
        - \sum_m \frac{
            \langle \Psi^N_0 | \hat{a}_s^{\dagger} \hat{a}_r^{\dagger} | \Psi^{N-2}_m \rangle 
            \langle \Psi^{N-2}_m | \hat{a}_p \hat{a}_q | \Psi^N_0 \rangle 
            }{
            \omega - \Omega^{N-2}_m - i\eta 
            }
\end{equation}
where $\hat{a}_p^{\dagger}$ and $\hat{a}_p$ are creation and annihilation operators in the second-quantization notation, 
$\Omega^{N\pm2}$ is the two-electron addition/removal energy, 
and $\eta$ is a positive infinitesimal number.
In Eq.~\ref{eq:pairing_matrix} and the following, 
we use $i$, $j$, $k$, $l$ for occupied orbitals, 
$a$, $b$, $c$, $d$ for virtual orbitals, 
$p$, $q$, $r$, $s$ for general molecular orbitals, 
and $m$ for the index of the two-electron addition/removal energy.
In ppRPA, 
$K$ is approximated in terms of the non-interacting $K^0$ via the Dyson equation\cite{vanaggelenExchangecorrelationEnergyPairing2013,vanaggelenExchangecorrelationEnergyPairing2014}
\begin{equation}\label{eq:dyson}
    K = K^0 + K^0 V K
\end{equation}

Eq.~\ref{eq:dyson} can be cast into a generalized eigenvalue problem that gives two-electron addition and removal energies\cite{vanaggelenExchangecorrelationEnergyPairing2013,yangExcitationEnergiesParticleparticle2014}
\begin{equation}\label{eq:eigen_equation}
\begin{bmatrix}\mathbf{A} & \mathbf{B}\\
\mathbf{B}^{\text{T}} & \mathbf{C}
\end{bmatrix}\begin{bmatrix}\mathbf{X}\\
\mathbf{Y}
\end{bmatrix}=\Omega\begin{bmatrix}\mathbf{I} & \mathbf{0}\\
\mathbf{0} & \mathbf{-I}
\end{bmatrix}\begin{bmatrix}\mathbf{X}\\
\mathbf{Y}
\end{bmatrix}
\end{equation}
with
\begin{align}
A_{ab,cd} & =\delta_{ac}\delta_{bd}(\epsilon_{a}+\epsilon_{b})+\langle ab||cd\rangle \label{eq:a_matrix} \\
B_{ab,kl} & =\langle ab||kl\rangle \\
C_{ij,kl} & =-\delta_{ik}\delta_{jl}(\epsilon_{i}+\epsilon_{j})+\langle ij||kl\rangle 
\end{align}
where $a<b$, $c<d$, $i<j$, $k<l$, 
$\Omega$ is the two-electron addition/removal energy,
$X$ and $Y$ are the two-electron addition and removal eigenvectors,
and the two-electron integral is defined as $\langle pq || rs \rangle = \langle pq | rs \rangle - \langle pq | sr \rangle$ with $\langle pq | rs \rangle = \int d\mathbf{x} d\mathbf{x'} \frac{\psi^*_p(\mathbf{x}) \psi_r(\mathbf{x}) \psi^*_q(\mathbf{x'}) \psi_s(\mathbf{x'})}{|\mathbf{r}-\mathbf{r'}|}$.
The ppRPA eigenvector is normalized as\cite{vanaggelenExchangecorrelationEnergyPairing2013,vanaggelenExchangecorrelationEnergyPairing2014}
\begin{equation} \label{eq:xy_norm}
    X^{m,\dagger} X^m - Y^{m,\dagger} Y^m = \pm 1
\end{equation}
where the upper sign is for two-electron addition excitations and the lower sign is for two-electron removal excitations.

In this work,
to obtain the neutral excitation energies of the $N$-electron system,
we first perform the self-consistent field (SCF) calculation of the ($N-2$)-electron system,
then calculate excitation energies as the differences between the lowest and a higher two-electron addition energies from Eq.~\ref{eq:eigen_equation}.
The neutral excitation energies of the $N$-electron system can also be obtained as the differences between the lowest and a higher two-electron removal energies of the ($N+2$)-electron system,
which has been used previously for predicting excitation energies of molecules and point defects\cite{bannwarthHoleHoleTamm2020,yuInitioNonadiabaticMolecular2020,hohensteinPredictionsPreedgeFeatures2021,liAccurateExcitationEnergies2024}.

For the studied point defects in this work,
the corresponding ($N-2$)-electron systems are closed-shell and can be calculated with spin-restricted DFT,
which is cheaper than spin-unrestricted DFT/HF typically used in TDDFT and quantum embedding approaches and is free of spin contamination.
For closed-shell systems,
the ppRPA equation in Eq.~\ref{eq:eigen_equation} can be cast into the spin-adapted equation\cite{yangBenchmarkTestsSpin2013},
where for triplet excitations, the ppRPA matrix elements are
\begin{align}
A^{\text{t}}_{ab,cd} & =\delta_{ac}\delta_{bd}(\epsilon_{a}+\epsilon_{b})+\langle ab||cd\rangle \label{eq:a_triplet}\\
B^{\text{t}}_{ab,kl} & =\langle ab||kl\rangle \label{eq:b_triplet}\\
C^{\text{t}}_{ij,kl} & =-\delta_{ik}\delta_{jl}(\epsilon_{i}+\epsilon_{j})+\langle ij||kl\rangle \label{eq:c_triplet} 
\end{align}
with  $a<b$, $c<d$, $i<j$ and $k<l$,
and for singlet excitations the ppRPA matrix elements are
\begin{align}
A^{\text{s}}_{ab,cd} & =\delta_{ac}\delta_{bd}(\epsilon_{a}+\epsilon_{b}) 
+ \frac{1}{\sqrt{(1+\delta_{ab})(1+\delta_{cd})}} (\langle ab|cd\rangle + \langle ab|dc\rangle) \label{eq:a_singlet}\\
B^{\text{s}}_{ab,kl} & = \frac{1}{\sqrt{(1+\delta_{ab})(1+\delta_{kl})}} 
(\langle ab|kl\rangle + \langle ab|lk\rangle) \label{eq:b_singlet}\\
C^{\text{s}}_{ij,kl} & =-\delta_{ik}\delta_{jl}(\epsilon_{i}+\epsilon_{j})
+ \frac{1}{\sqrt{(1+\delta_{ij})(1+\delta_{kl})}} (\langle ij|kl\rangle + \langle ij|lk\rangle) \label{eq:c_singlet}
\end{align}
with  $a \leq b$, $c \leq d$, $i \leq j$ and $k \leq l$.

For point defects whose corresponding ($N-2$)-electron system face convergence difficulties in SCF calculations,
KS orbitals and orbital energies obtained from the $N$-electron system are used in Eq.~\ref{eq:eigen_equation},
which is denoted as ppRPA*\cite{yangDoubleRydbergCharge2013}.
In ppRPA*,
the reference is a non-optimized single determinant for the ($N-2$)-electron system, 
similar in spirit to spin-flip methods.
ppRPA* has been applied to predict valence, Rydberg, and double excitation energies for small molecules\cite{yangDoubleRydbergCharge2013}, 
and we will investigate its reliability for defect systems here.

The ppRPA working equation defined in Eq.~\ref{eq:eigen_equation} has a similar structure as the Casida equation in TDDFT\cite{casidaTimeDependentDensityFunctional1995,ullrichTimeDependentDensityFunctionalTheory2011},
which can be solved with the Davidson algorithm with the $\mathcal{O}(N^4)$ scaling ($N$ is the number of orbitals in the system)\cite{yangExcitationEnergiesParticleparticle2014}.
In this work,
the Davidson algorithm is combined with the active space approach developed in Ref.~\citenum{liLinearScalingCalculations2023} to lower the cost of ppRPA calculations.
The full-space results are obtained through an extrapolation scheme using data from a series of different active-space calculations.

\subsection{Natural Transition Orbital in ppRPA}

To analyze the character of defect excited states,
we further develop the NTO approach within ppRPA.
The NTO concept has been widely used in particle-hole formalisms such as TDDFT and BSE to provide qualitative descriptions of electronic transitions\cite{martinNaturalTransitionOrbitals2003,choSimplifiedGWBSE2022}.
In the NTO approach of particle-hole formalisms,
the dominant particle-hole pair in the excited state is obtained by the SVD of the corresponding transition density matrix\cite{martinNaturalTransitionOrbitals2003}.

Parallel to NTOs in the particle-hole formalisms that describe particle-hole transitions,
NTOs in ppRPA convey information about particle-particle pairs and hole-hole pairs,
which are obtained from the SVD of the two-electron addition and removal eigenvectors separately
\begin{align}
    X^m = & C^{\text{p1}, m} \sqrt{\lambda^{\text{p}, m}} C^{\text{p2}, m\dagger} \label{eq:nto_pp} \\
    Y^m = & C^{\text{h1}, m} \sqrt{\lambda^{\text{h}, m}} C^{\text{h2}, m\dagger} \label{eq:nto_hh}
\end{align}
In Eq.~\ref{eq:nto_pp},
the NTO coefficients $C^{\text{p1}}$ and $C^{\text{p2}}$ weighted with $\lambda^{\text{p}}$ are associated with particle-particle pairs for adding the first and second electrons.
Similarly,
in Eq.~\ref{eq:nto_hh},
the NTO coefficients $C^{\text{h1}}$ and $C^{\text{h2}}$ weighted with $\lambda^{\text{h}}$ are associated with hole-hole pairs for removing the first and second electrons.
As a consequence of the normalization in Eq.~\ref{eq:xy_norm},
the NTO weights satisfy the following relation
\begin{equation}\label{eq:nto_weight_constrain}
    \sum_a^{\text{vir}} \lambda^{\text{p}}_a - \sum_i^{\text{occ}} \lambda^{\text{h}}_i = \pm 1
\end{equation}
where the upper sign is for two-electron addition excitations and the lower sign is for two-electron removal excitations. 
The resulting NTO weights ($\lambda_a^{\text{p}}$ and $\lambda_i^{\text{h}}$) can thus be employed to qualitatively analyze the components and multireference character of the associated ground and excited states.

\section{COMPUTATIONAL DETAILS}\label{sec:comp_detail}

Ground-state geometries of all three defect systems were optimized with the PBE functional~\cite{perdewGeneralizedGradientApproximation1996} using the Quantum ESPRESSO package~\cite{giannozziQuantumESPRESSOExascale2020,carnimeoQuantumESPRESSOOne2023}, 
and details can be found in the Supporting Information (SI). 
We then performed ($N-2$)-electron ground-state DFT calculations for VC in diamond and VO in MgO.
For C$_{\text{B}}$C$_{\text{N}}$ in two-dimensional h-BN,
due to the convergence difficulty in the ($N-2$)-electron ground-state DFT calculation, 
the $N$-electron ground-state DFT calculation was performed.
All ground-state DFT calculations in periodic Gaussian basis sets were carried out using the PySCF quantum chemistry software package~\cite{sunPySCFPythonbasedSimulations2018,sunRecentDevelopmentsPySCF2020} 
with $\Gamma$-point sampling and Gaussian density fitting. 
Two functionals (PBE\cite{perdewGeneralizedGradientApproximation1996} and B3LYP~\cite{beckeDensityFunctionalThermochemistry1993,leeDevelopmentColleSalvettiCorrelationenergy1988}) were used in combination with the cc-pVDZ basis set~\cite{dunningGaussianBasisSets1989} and the corresponding cc-pVDZ-RI auxiliary basis set~\cite{weigendEfficientUseCorrelation2002}. 
With electron integrals and DFT solutions obtained from PySCF, 
we further performed active-space ppRPA calculations with periodic boundary condition to predict VEEs of point defects using the \texttt{\detokenize{Lib_ppRPA}} library\cite{liUnpublished}. 
As shown in the SI,
using the cc-pVTZ basis set leads to very close VEE results as those obtained with the cc-pVDZ basis set, 
so we only present results computed using the cc-pVDZ basis.

\section{RESULTS}\label{sec:results}

\subsection{Extrapolation of Active-Space Results to Full-Space Limit}\label{subsec:extrapolation}

\begin{figure}
\includegraphics[width=0.75\textwidth]{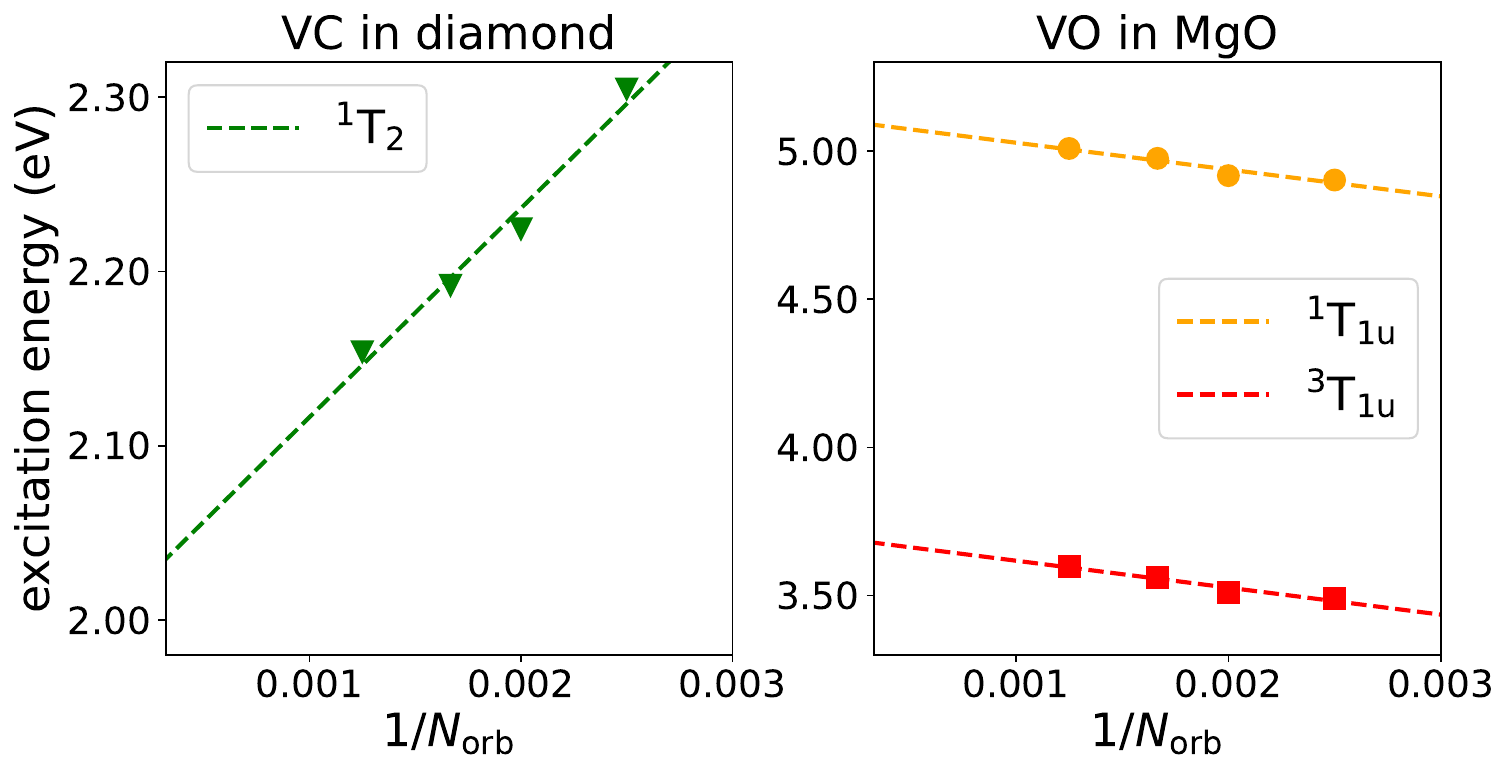}
\caption{Extrapolations of VEEs with respect to the inverse of the active-space size.
$N_{\text{orb}}$ is the number of canonical orbitals in the active space, and the cc-pVDZ basis set was used.
Left: $^1$T$_2$ state of VC in diamond with D$_{2\text{d}}$ symmetry (215-atom supercell).
Right: $^1$T$_{1\text{u}}$ and $^3$T$_{1\text{u}}$ states of VO in MgO (215-atom supercell).}
\label{fig:extrapolation}
\end{figure}

We first establish the extrapolation scheme to obtain the full-space results in the active-space ppRPA approach.
The VEEs of VC in diamond with D$_{2\text{d}}$ symmetry and VO in MgO obtained from ppRPA@B3LYP using supercell models containing 215 atoms are shown in Fig.~\ref{fig:extrapolation}.
The supercell of VC in diamond has 644 occupied and 2366 virtual orbitals,
while the supercell of VO in MgO has 1075 occupied and 2043 virtual orbitals.
As shown in Eq.~\ref{eq:a_matrix}, 
the $\mathbf{A}$ matrix in ppRPA has four indexes of virtual orbitals,
so an active-space approach is desired to reduce the computational cost in the calculations of large supercell models.
For simplicity,
the active spaces used in this work include the same numbers of occupied and virtual canonical molecular orbitals around the Fermi level,
which are consistent with Refs.~\citenum{liLinearScalingCalculations2023} and~\citenum{liAccurateExcitationEnergies2024}.
As shown in Fig.~\ref{fig:extrapolation},
similar convergence patterns are observed for both point defect systems.
We find a linear relationship between the excitation energies and the inverse of the number of active-space orbitals (denoted as $N_{\text{orb}}$).
The excitation energy at the full-space limit is thus obtained from the extrapolation of active-space values against $1/N_{\text{orb}}$, 
similar to schemes adopted in the quantum embedding literature~\cite{lauOpticalPropertiesDefects2024,haldarLocalExcitationsCharged2023}.
In this work, we use the active spaces containing 400, 500, 600, and 800 orbitals in the four-point extrapolation scheme to obtain the full-space results. 

Compared with previous works for molecular~\cite{liLinearScalingCalculations2023}
and point defect systems~\cite{liAccurateExcitationEnergies2024},
the convergence of active-space ppRPA results is slower in this study for two reasons.
First,
the number of orbitals needed in the active space for periodic systems is larger due to the denser manifold of low-lying states in solids than in molecules.
Second,
the convergence of two-electron addition energies from the particle-particle channel mainly depends on the virtual space,
which is more difficult to converge than the occupied space in the hole-hole channel~\cite{liAccurateExcitationEnergies2024}.
Nevertheless,
the size of the active space needed for reliable extrapolation is still much smaller than the size of the full space.
We note, in principle, 
natural orbitals from diagonalizing the density matrix of a correlated or excited-state method can be used to construct the active space and further reduce the computational cost.

\subsection{Vertical Excitation Energies}\label{subsec:vee}

\subsubsection{VC in diamond}

\FloatBarrier

\begin{figure}
\includegraphics[width=0.6\textwidth]{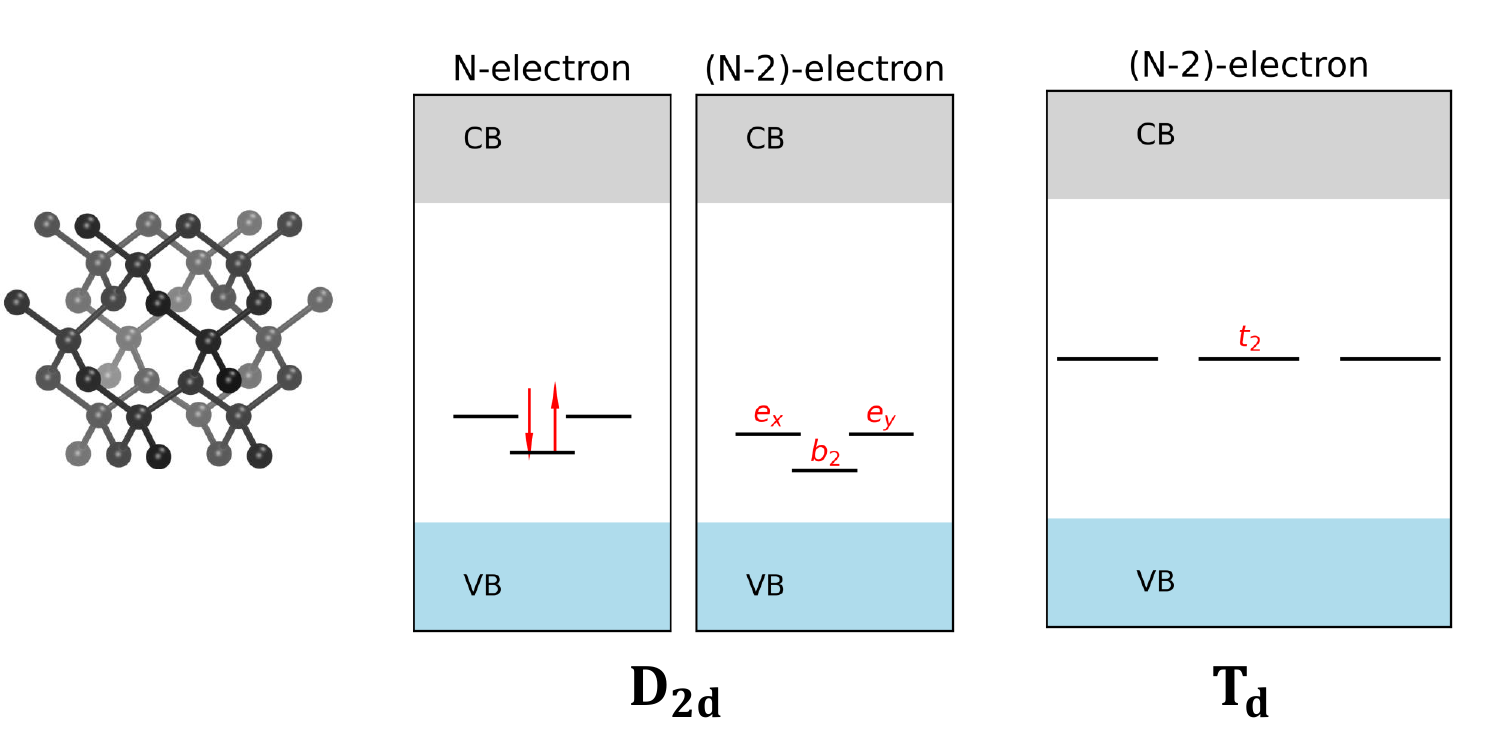}
\caption{Illustration of defect energy levels and ground-state electron configurations of VC in diamond with D$_{2\text{d}}$ and T$_{\text{d}}$ symmetries (energy levels are qualitative only). 
The electron configuration of $N$-electron state with T$_{\text{d}}$ symmetry is not shown because it is a multiconfigurational state.}
\label{fig:diamond}
\end{figure}

\begin{table}[!ht]
\renewcommand{\arraystretch}{1.2}
\setlength\tabcolsep{15pt}
\caption{VEEs of VC in diamond obtained from the ppRPA approach based on PBE and B3LYP functionals compared with reference values.
Supercell-size extrapolated values were obtained from the results of supercells containing $63$ and $215$ atoms. The geometry with D$_{\text{2d}}$ or T$_{\text{d}}$ symmetry was employed, where the corresponding excited state is $^1$E (D$_{\text{2d}}$) or $^1$T$_2$ (T$_{\text{d}}$) state.
The cc-pVDZ basis set was used. 
All values are in \,{eV}.
}\label{tab:diamond}
\begin{tabular}{c|c|c}
\hline
                                                                              Method&Structure & $^1$E/$^1$T$_2$     \\
\hline      
Experiment\cite{lannooOpticalAbsorptionNeutral1968}                          & & 2.2           \\
ppRPA@PBE (supercell 215)                                   & D$_{\text{2d}}$ & 1.67          \\
ppRPA@PBE (extrapolated)                                  &  D$_{\text{2d}}$  & 1.77          \\
ppRPA@B3LYP (supercell 215)                               &  D$_{\text{2d}}$  & 2.02          \\
ppRPA@B3LYP (extrapolated)                                &  D$_{\text{2d}}$  & 2.15          \\
ppRPA@PBE (supercell 215)                                  & T$_{\text{d}}$   & 1.51          \\
ppRPA@PBE (extrapolated)                                   &  T$_{\text{d}}$  & 1.56          \\
ppRPA@B3LYP (supercell 215)                                &  T$_{\text{d}}$  & 1.81          \\
ppRPA@B3LYP (extrapolated)                                 &  T$_{\text{d}}$  & 1.89          \\
\hline         
TDDFT@PBE (supercell 215)                                 &  D$_{\text{2d}}$   & 1.19          \\
TDDFT@PBE (extrapolated)                                  &  D$_{\text{2d}}$   & 1.26          \\
TDDFT@B3LYP (supercell 215)                               &  D$_{\text{2d}}$   & 1.30          \\
TDDFT@B3LYP (extrapolated)                                &  D$_{\text{2d}}$   & 1.40          \\
CCSD\cite{lauOpticalPropertiesDefects2024}              &  D$_{\text{2d}}$  & 2.09          \\
DMC\cite{hoodQuantumMonteCarlo2003}                     &   T$_{\text{d}}$  & 1.51$\pm$0.34 \\
$\Delta$SCF@B3LYP\cite{mackrodtCalculatedEnergiesCharge2022} & T$_{\text{d}}$ & 1.57          \\
\hline
\end{tabular}
\end{table}

Our first tested system is VC in diamond.
The predicted VEEs of its excited state from the ppRPA approach based on PBE and B3LYP are presented in Table~\ref{tab:diamond}. 
To correct the finite supercell-size error, 
we also employed a two-point supercell-size extrapolation scheme, 
using 63-atom and 215-atom excitation energies in a linear fitting of the form: 
$E(1/N_\mathrm{atom}) = E_\infty + a/N_\mathrm{atom}$,
which has been successfully used to predict VEEs of defect systems in the thermodynamic limit~\cite{jinExcitedStateProperties2023,vermaOpticalPropertiesNeutral2023,liAccurateExcitationEnergies2024}.
Ground-state geometries of both D$_{\text{2d}}$ and T$_{\text{d}}$ symmetries were used in ppRPA calculations. In the calculations with the D$_{\text{2d}}$ geometry,
the symmetry of the studied defect excited state is $^1$E, while the defect excited state is $^1$T$_2$ in the T$_{\text{d}}$ geometry.

As shown in Fig.~\ref{fig:diamond},
in the statically distorted D$_{\text{2d}}$ geometry,
the lowest defect orbital ($b_2$) is doubly occupied and the unoccupied defect orbitals ($e_x$ and $e_y$) are two-fold degenerate.
In the corresponding ($N-2$)-electron state, two electrons are removed from the lowest defect orbital ($b_2$).
Thus, both $N$-electron and ($N-2$)-electron ground states may be described by a single determinant.
Although the $N$-electron ground state is a closed-shell singlet, 
TDDFT based on both GGA and hybrid functionals (i.e., PBE and B3LYP) has large errors around or larger than $0.8$ \,{eV} for predicting the VEE.
This large error can be attributed to the insufficient description of the ground state using a single KS determinant.
As will be seen in Section~\ref{subsec:multi},
the ppRPA@B3LYP NTO analysis shows that the singlet ground state in the D$_{\text{2d}}$ geometry consists of $\sim87\%$ contribution from the $|b_2 \bar{b}_2 \rangle$ configuration,
while the remaining $\sim13\%$ contribution comes from the doubly-excited configurations $|e_x \bar{e}_x \rangle$ and $|e_y \bar{e}_y \rangle$.
Single-determinant TDDFT starts with a pure $|b_2 \bar{b}_2 \rangle$ configuration,
which is responsible for its limited accuracy.
In ppRPA, these two defect-level electrons are treated in a subspace configuration interaction (CI) fashion,
so near-degenerate configurations can be treated on equal footing.
ppRPA based on PBE underestimates the VEE by more than $0.4$ \,{eV},
which agrees with our previous results for other defect systems\cite{liAccurateExcitationEnergies2024}, 
although it is already substantially better than TDDFT.
The extrapolated ppRPA@B3LYP provides further improved accuracy.
It predicts accurate VEE with an error smaller than $0.1$ \,{eV},
which is comparable to more expensive EOM-CCSD (equation-of-motion coupled-cluster singles and doubles)~\cite{lauOpticalPropertiesDefects2024}.

Next, we turn to the results based on the T$_{\text{d}}$ geometry,
which is consistent with the tetrahedral symmetry observed in the experiment\cite{lannooOpticalAbsorptionNeutral1968}
(although the D$_{\text{2d}}$ geometry may also be relevant through a dynamic Jahn-Teller mechanism\cite{lannooOpticalAbsorptionNeutral1968,clarkNeutralVacancyDiamond1997,daviesJahnTellerEffectVibronic1981}).
As shown in Fig.~\ref{fig:diamond},
orbitals with $t_2$ symmetry are threefold degenerate.
For the $N$-electron system,
two electrons are filled into three degenerate orbitals,
indicating that the ground state has a strong multiconfigurational character.
As a result,
single-reference approaches are not appropriate here.
By removing two electrons,
the ground state of the ($N-2$)-electron system used in ppRPA is a well-behaved closed-shell singlet state without dramatic static correlation,
which can be well described by DFT.
Therefore,
ppRPA does not suffer from large static correlation error and predicts accurate VEEs.
%In ppRPA calculations,
%the $^3$T$_1$ state is predicted as the ground state.
%Thus,
%the VEE of the $^1$T$_2$ state is obtained as the difference between two-electron addition energies of $^1$T$_2$ and $^3$T$_1$.
To directly compare with the literature values,
the VEE of the $^1$T$_2$ state is obtained as the difference between two-electron addition energies of $^1$T$_2$ excited and $^1$E ground state.
As shown in Table~\ref{tab:diamond},
both ppRPA@PBE and ppRPA@B3LYP provide reasonably accurate results; for example, ppRPA@B3LYP only underestimates the VEE by 0.3 eV. This performance is more accurate than $\Delta$SCF and even slightly better than the more costly diffusion quantum Monte Carlo (DMC).
% TODO: why Td results underestimate
Through a seamless combination of DFT for the ($N-2$)-electron system and a subspace CI for the two valence electrons, 
ppRPA provides a balanced description of the challenging ground and excited states in the T\textsubscript{d} geometry with a low cost. We note that, the slightly larger errors in ppRPA predictions for the T\textsubscript{d} geometry may result from less accurate DFT geometry optimization, where each of the three degenerate $t_2$ KS orbitals was enforced to accommodate $2/3$ of an electron.

\FloatBarrier

\subsubsection{VO in MgO}

\FloatBarrier

\begin{figure}
\includegraphics[width=0.6\textwidth]{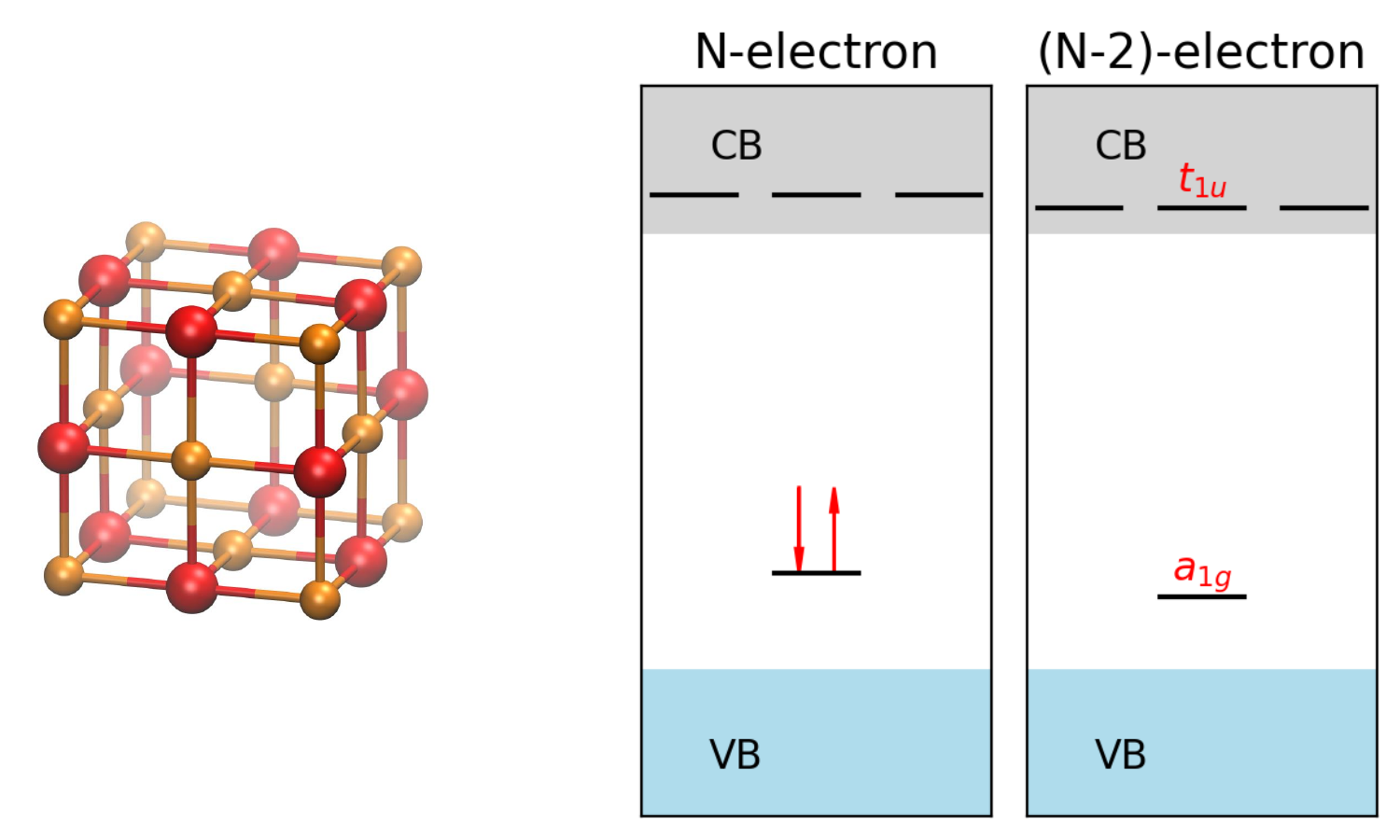}
\caption{Illustration of defect energy levels and ground-state electron configurations of VO in MgO (energy levels are qualitative only).}
\label{fig:mgo}
\end{figure}

\begin{table}[!ht]
\renewcommand{\arraystretch}{1.2}
\setlength\tabcolsep{15pt}
\caption{VEEs of VO in MgO obtained from the ppRPA approach based on PBE and B3LYP functionals compared with reference values.
Supercell-size extrapolated values were obtained from the results of supercells containing $63$ and $215$ atoms.
The geometry of the $^1$A$_{1\text{g}}$ ground state and cc-pVDZ basis set were used. 
All values are in \,{eV}.
}\label{tab:mgo}
\begin{tabular}{c|cc}
\hline
                                                                                   Method & $^1$T$_{1\text{u}}$ & $^3$T$_{1\text{u}}$ \\
\hline
Experiment\cite{chenDefectClusterCenters1969,kappersEnsuremathCentersMagnesium1970} & 5.0                 & 2.3$\sim$2.4 (emission)       \\
ppRPA@PBE (supercell 215)                                                           & 4.89                & 3.67                \\
ppRPA@PBE (extrapolated)                                                            & 4.82                & 3.57                \\
ppRPA@B3LYP (supercell 215)                                                         & 5.10                & 3.69                \\
ppRPA@B3LYP (extrapolated)                                                          & 5.05                & 3.58                \\
\hline                       
TDDFT@PBE (supercell 215)                                                           & 3.86                & 3.29                \\
TDDFT@PBE (extrapolated)                                                            & 3.37                & 3.08                \\
TDDFT@B3LYP (supercell 215)                                                         & 4.60                & 3.57                \\
TDDFT@B3LYP (extrapolated)                                                          & 4.29                & 3.42                \\
BSE/$G_0W_0$@LDA\cite{rinkeFirstPrinciplesOpticalSpectra2012}                       &                     & 3.40                \\
CCSD\cite{lauOpticalPropertiesDefects2024}                                          & 5.31                &                     \\
CCSD\cite{galloPeriodicEquationofmotionCoupledcluster2021}                          & 5.28                & 3.66                \\
CASPT2\cite{sousaAccuratePredictionOptical2001}                                     & 5.44                & 4.09                \\
CAS-DMET\cite{vermaOpticalPropertiesNeutral2023}                                    & 6.26                & 2.74                \\
NEVPT2-DMET\cite{vermaOpticalPropertiesNeutral2023}                                 & 5.24                & 2.89                \\
embedded-BSE@DDH\cite{vorwerkDisentanglingPhotoexcitationPhotoluminescence2023}     & 5.23                & 2.93                \\
FN-DMC\cite{ertekinPointdefectOpticalTransitions2013}                               &                     & 3.80                \\
\hline
\end{tabular}
\end{table}

We now turn to the discussion of VO in MgO.
The VEEs of the $^1$T$_{1\text{u}}$ and $^3$T$_{1\text{u}}$ states of VO in MgO obtained from ppRPA based on PBE and B3LYP using the 215-atom supercell and supercell-size extrapolation scheme are presented in Table~\ref{tab:mgo}. 
As shown in Fig.~\ref{fig:mgo},
the $s$-type $a_{1g}$ orbital is within the band gap and three $p$-type $t_{1u}$ orbitals with higher energies are in the conduction band.
Both $N$-electron and ($N-2$)-electron ground states can be well described by a single KS determinant.
We find that TDDFT based on PBE and B3LYP largely underestimates the excitation energy of the $^1$T$_{1\text{u}}$ state by $0.7\sim1.6$ \,{eV}.
As shown in Ref.~\citenum{jinExcitedStateProperties2023},
functionals with a higher percentage of the HF exchange are needed for TDDFT to yield improved results.
In addition to large underestimations,
TDDFT shows an undesired starting-point dependence for VO in MgO, 
where the difference in VEEs of the $^1$T$_{1\text{u}}$ state from PBE vs. B3LYP is $0.9$ \,{eV}.
Embedding approaches, including regional embedding with the EOM-CCSD solver, NEVPT2-DMET, and embedded BSE@DDH, provide significantly improved accuracy over TDDFT based on conventional functionals,
which slightly overestimate the VEE of the $^1$T$_{1\text{u}}$ state by $0.2\sim0.3$ \,{eV}.

Our ppRPA approach based on both PBE and B3LYP achieves excellent performance.
The error of ppRPA@B3LYP for predicting the VEE of the $^1$T$_{1\text{u}}$ state is only $0.05$ \,{eV} compared with the experiment value.
In addition,
ppRPA has a weaker DFT starting-point dependence than TDDFT.
The difference between VEEs of the $^1$T$_{1\text{u}}$ state obtained from ppRPA@PBE and ppRPA@B3LYP is only around $0.2$ \,{eV},
much smaller than the $0.9$ \,{eV} difference between TDDFT@PBE and TDDFT@B3LYP.

For the VEE of the $^3$T$_{1\text{u}}$ state,
all methods in Table~\ref{tab:mgo} overestimate compared with the experimental value 
since the experimental measurement was conducted for the emission process. 
As suggested in Ref.~\citenum{jinExcitedStateProperties2023},
an estimation of the Franck-Condon shift is needed for a direct comparison with the emission peak.
Nevertheless, comparing to higher-level theories, our ppRPA predictions of the $^3$T$_{1\text{u}}$ VEE are less than 0.1 eV different from CCSD~\cite{lauOpticalPropertiesDefects2024} and around 0.2 eV different from fixed-node DMC (FN-DMC)~\cite{ertekinPointdefectOpticalTransitions2013}. 
Also, the functional dependence in ppRPA (PBE vs. B3LYP) is only $0.02$ \,{eV} for the $^3$T$_{1\text{u}}$ state.

\FloatBarrier

\subsubsection{C$_{\text{B}}$C$_{\text{N}}$ in h-BN}

\FloatBarrier

\begin{figure}
\includegraphics[width=0.6\textwidth]{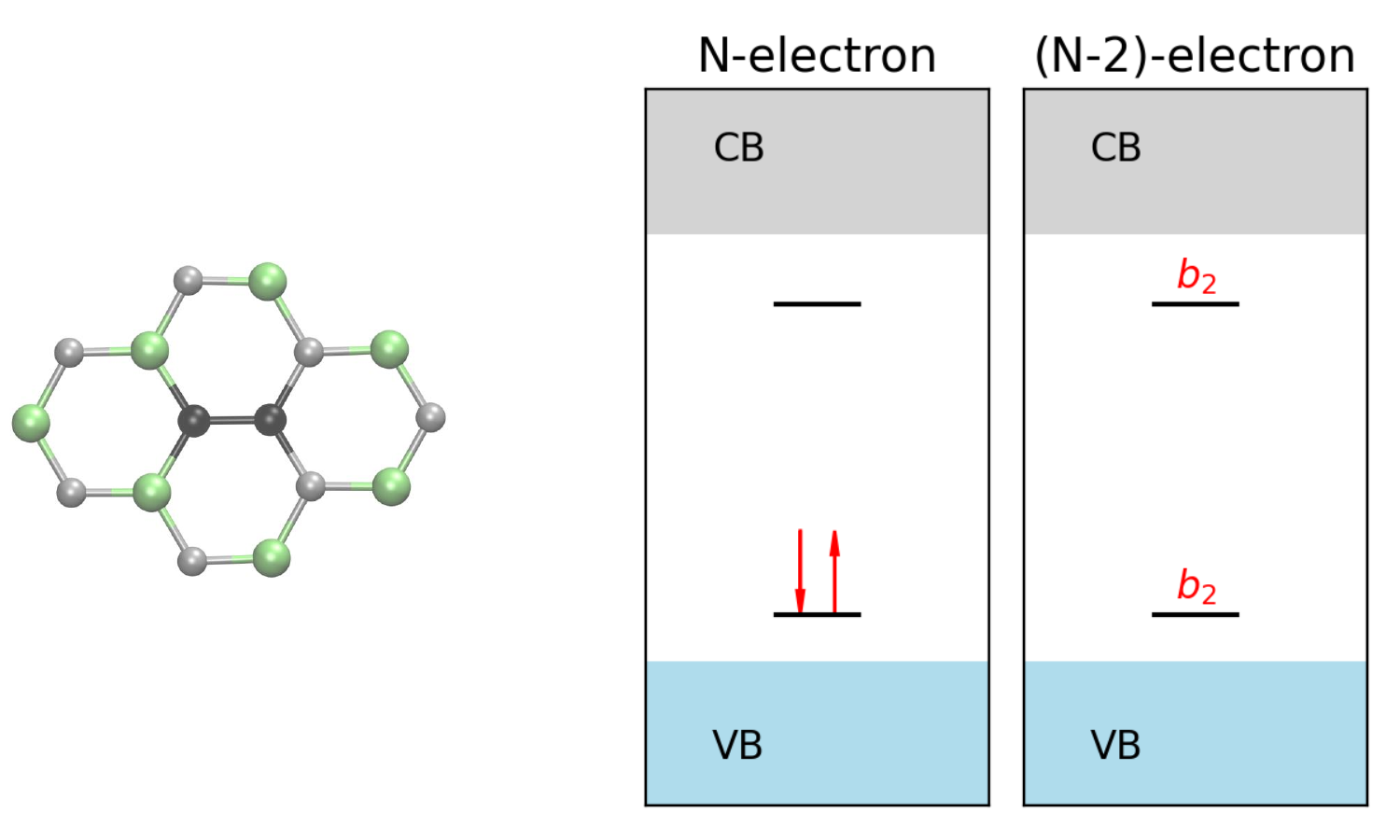}
\caption{Illustration of defect energy levels and ground-state electron configurations of C$_{\text{B}}$C$_{\text{N}}$ in h-BN (energy levels are qualitative only).}
\label{fig:cbcn}
\end{figure}

\begin{table}[!ht]
\renewcommand{\arraystretch}{1.2}
\setlength\tabcolsep{15pt}
\caption{VEEs of C$_{\text{B}}$C$_{\text{N}}$ in h-BN obtained from the ppRPA* approach based on PBE and B3LYP functionals compared with reference values. 
The geometry of the $^1$A$_1$ ground state and cc-pVDZ basis set were used. 
All values are in \,{eV}.
}\label{tab:bn}
\begin{tabular}{c|c}
\hline
                                                                      Method & $^1$A$_1$ \\
\hline  
Experiment\cite{winterPhotoluminescentPropertiesCarbondimer2021} \textsuperscript{\emph{a}}     & 4.6       \\
ppRPA*@PBE (supercell 128)                                            & 3.78      \\
ppRPA*@B3LYP (supercell 128)                                          & 4.54      \\
\hline
TDDFT@PBE (supercell 128)                                             & 4.03      \\
TDDFT@B3LYP (supercell 128)                                           & 4.63      \\
TDDFT@CAM-B3LYP\cite{koronaExploringPointDefects2019}                 & 4.78      \\
TDDFT@PBE0\cite{winterPhotoluminescentPropertiesCarbondimer2021}      & 4.61      \\
BSE/ev$GW$@PBE0\cite{winterPhotoluminescentPropertiesCarbondimer2021} & 4.64      \\
$\Delta$SCF@HSE($\alpha=0.40$)\cite{mackoit-sinkevicieneCarbonDimerDefect2019} \textsuperscript{\emph{b}}         & 4.53      \\
CCSD\cite{lauOpticalPropertiesDefects2024}                            & 4.76      \\
cRPA@PBE\cite{muechlerQuantumEmbeddingMethods2022} \textsuperscript{\emph{b}}                  & 3.98      \\
cRPA@HSE\cite{muechlerQuantumEmbeddingMethods2022} \textsuperscript{\emph{b}}                      & 4.23  \\
\hline
\end{tabular}

\textsuperscript{\emph{a}} Estimated vertical excitation energy for C$_\text{B}$C$_\text{N}$ in two-dimensional h-BN.

\textsuperscript{\emph{b}} Calculations were performed for C$_\text{B}$C$_\text{N}$ in bulk-layered h-BN.
\end{table}

Our final tested system is C$_{\text{B}}$C$_{\text{N}}$ in two-dimensional h-BN.
The VEEs of the $^1$A$_1$ state of C$_{\text{B}}$C$_{\text{N}}$ in h-BN obtained from the ppRPA* approach based on PBE and B3LYP using the 128-atom supercell are presented in Table~\ref{tab:bn}. 
Due to the difficulty in the SCF convergence of the corresponding ($N-2$)-electron system of C$_{\text{B}}$C$_{\text{N}}$ in h-BN,
KS orbitals and orbital energies from the $N$-electron system are used in ppRPA* calculations.
As shown in Ref.~\citenum{lauOpticalPropertiesDefects2024},
converged excitation energies of C$_{\text{B}}$C$_{\text{N}}$ in h-BN can be obtained using the 128-atom supercell model.
%Both TDDFT and $\Delta$SCF based on hybrid functionals give VEE errors smaller than $0.2$ \,{eV},
TDDFT based on hybrid functionals gives VEE errors smaller than $0.2$ \,{eV},
which are comparable to the results of BSE/ev$GW$@PBE0 and EOM-CCSD.
The slightly larger errors in cRPA-based methods may stem from the quality of the approximated dynamical correlation and the use of three-dimensional h-BN structure
\cite{lauOpticalPropertiesDefects2024}.
ppRPA*@PBE severely underestimates the VEE by $0.8$ \,{eV}, 
which is similar to the result of TDDFT@PBE. 
We suspect the poor performance of ppRPA*@PBE is due to the use of unrelaxed KS orbitals in the non-optimized $(N-2)$-electron ground state.
ppRPA* based on B3LYP provides significantly improved accuracy with a VEE error smaller than $0.1$ \,{eV}.
As shown in the SI,
the VEE of C$_{\text{B}}$C$_{\text{N}}$ in h-BN obtained from ppRPA* also converges rapidly with respect to the size of the active space. However, we note that the large functional dependence in ppRPA* suggests the importance of orbital optimization in the $(N-2)$-electron DFT ground state.

\FloatBarrier

\subsection{Analysis of Multireference Character}\label{subsec:multi}

\FloatBarrier

\begin{table}[!ht]
\renewcommand{\arraystretch}{1.2}
\setlength\tabcolsep{15pt}
\caption{NTO weights and associated electron configurations of different defect states obtained from ppRPA@B3LYP.
NTO weights larger than $0.05$ are shown.
}\label{tab:nto}
\begin{tabular}{cc|c}
\hline
     System              & State   & NTO weight \\
\hline
NV$^-$ in diamond  & $^3$A$_2$  
                   & 0.990 $|a_1 \bar{a}_1 e_x e_y\rangle$ \\
                   & $^1$A$_1$  
                   & 0.424 $|a_1 \bar{a}_1 e_x \bar{e}_x\rangle$, 
                     0.423 $|a_1 \bar{a}_1 e_y \bar{e}_y\rangle$, 
                     0.152 $|e_x \bar{e}_x e_y \bar{e}_y\rangle$ \\
SiV$^0$ in diamond & $^3$A$_{2\text{g}}$ 
                   & 0.906 $|e_{ux} \bar{e}_{ux} e_{uy} \bar{e}_{uy} e_{gx} e_{gy}\rangle$ \\
                   & $^3$A$_{2\text{u}}$ 
                   & 0.475 $|e_{ux} \bar{e}_{ux} e_{uy} e_{gx} e_{gy} \bar{e}_{gy}\rangle$,
                     0.475 $|e_{ux} e_{uy} \bar{e}_{uy} e_{gx} \bar{e}_{gx} e_{gy}\rangle$ \\
                   & $^3$A$_{1\text{u}}$ & 
                     0.467 $|e_{ux} \bar{e}_{ux} e_{uy} e_{gx} \bar{e}_{gx} e_{gy}\rangle$,
                     0.467 $|e_{ux} e_{uy} \bar{e}_{uy} e_{gx} e_{gy} \bar{e}_{gy}\rangle$\\
VV$^0$ in 4H-SiC   & $^3$A$_2$ 
                   & 0.973 $|a_1 \bar{a}_1 e_x e_y\rangle$ \\
                   & $^1$A$_1$     
                   & 0.443 $|a_1 \bar{a}_1 e_x \bar{e}_x\rangle$, 
                     0.442 $|a_1 \bar{a}_1 e_y \bar{e}_y\rangle$, 
                     0.114 $|e_x \bar{e}_x e_y \bar{e}_y\rangle$  \\
VO in MgO          & $^1$A$_{1\text{g}}$ 
                   & 0.982 $|a_{1g} \bar{a}_{1g}\rangle$ \\
VC in diamond (D$_\mathrm{2d}$)      & $^1$A$_1$  
                   & 0.875 $|b_2 \bar{b}_2\rangle$, 
                     0.067 $|e_x \bar{e}_x\rangle$, 
                     0.067 $|e_y \bar{e}_y\rangle$ \\
\hline
\end{tabular}
\end{table}

\begin{figure}
\includegraphics[width=0.6\textwidth]{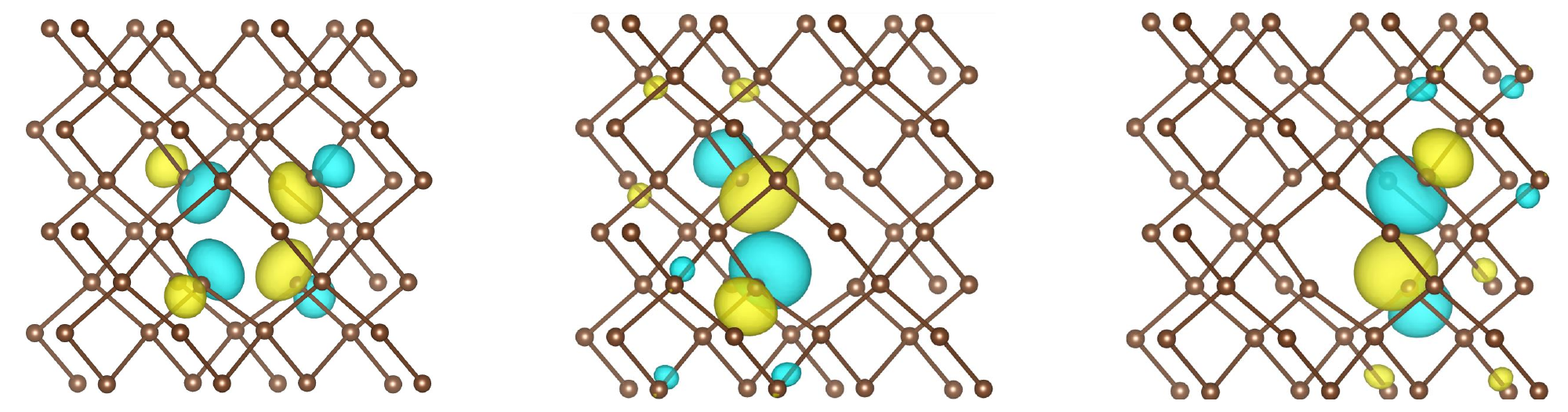}
\caption{Dominant two-electron addition NTOs of the $^1$A$_1$ ground state of VC in diamond (D$_{\text{2d}}$) obtained from ppRPA@B3LYP calculation of the 63-atom supercell. 
Isosurface value is 0.08 a.u. 
NTO weights are 0.875 (left), 0.067 (middle), and 0.067 (right).}
\label{fig:nto}
\end{figure}

In this section,
we further analyze the character of defect ground and excited states using the NTO approach in ppRPA on systems studied in this and previous works~\cite{liAccurateExcitationEnergies2024}.
NTO weights and associated electron configurations of different defect states obtained from ppRPA@B3LYP are tabulated in Table~\ref{tab:nto}.
%For simplicity, only non-degenerate defect states are discussed.
For NV$^-$ in diamond,
only one dominant NTO weight of $0.990$ can be found in the $^3$A$_2$ ground state,
which means it can be properly described by a single-determinant method.
For the $^1$A$_1$ state,
two approximately equivalent NTO weights corresponding to two singly-excited electron configurations indicate a strong multireference character.
The doubly-excited configuration $|e_x \bar{e}_x e_y \bar{e}_y\rangle$ with a NTO weight of $0.152$ is also found in the $^1$A$_1$ state,
which agrees with the analysis using QDET in Ref.~\citenum{jinVibrationallyResolvedOptical2022}.
Similar observations are found for VV$^0$ in 4H-SiC,
where the $^3$A$_2$ ground state is dominant with a single determinant, and the $^1$A$_1$ state has a strong multireference character.
For SiV$^0$ in diamond,
the $^3$A$_{2\text{g}}$ ground state shows a dominant NTO weight of $0.906$, 
while two excited states ($^3$A$_{2\text{u}}$ and $^3$A$_{1\text{u}}$) show strong multireference characters.  
The dominant NTO weight of $0.982$ can be found in the $^1$A$_{1\text{g}}$ ground state of VO in MgO,
which explains the good accuracy of single-reference methods in Table~\ref{tab:mgo}.
For VC in diamond of the D$_{\text{2d}}$ symmetry,
the multireference character is found in the $^1$A$_1$ ground state.
In addition to the $87\%$ contribution from the $|b_2 \bar{b}_2 \rangle$ configuration,
the singlet ground state has $13\%$ contribution from doubly-excited configurations $|e_x \bar{e}_x \rangle$ and $|e_y \bar{e}_y \rangle$,
which cannot be captured by the single-determinant KS-DFT approach. 
This analysis agrees with the unsatisfactory accuracy of TDDFT in Table~\ref{tab:diamond}.

Furthermore, in Fig.~\ref{fig:nto},
we plot the dominant two-electron addition NTOs of the $^1$A$_1$ ground state of VC in diamond (D$_{\text{2d}}$ geometry) obtained from ppRPA@B3LYP.
The electron configurations of this state can be approximately described by three doubly-occupied NTOs as $87.5\%$ contribution from $|b_2 \bar{b}_2\rangle$,
$6.7\%$ contribution from $|e_x \bar{e}_x\rangle$ and $6.7\%$ contribution from $|e_y \bar{e}_y\rangle$.
%%%%%%%%%%%%%%%%%%%%%===> Comment from JY <===%%%%%%%%%%%%%%%%%%%%%
% I don't think weights can be directly taken as the coefficients %
% used for the linear combination of wave functions.              %
%%%%%%%%%%%%%%%%%%%%%%%%%%%%%%%%%%%%%%%%%%%%%%%%%%%%%%%%%%%%%%%%%%%
These NTOs show that the two valence electrons are relatively localized around the carbon vacancy center.

\FloatBarrier

\section{CONCLUSIONS}\label{sec:conclusion}
In summary, 
we applied the ppRPA approach within the particle-particle channel to predict accurate excitation energies of point defect systems in this work.
In ppRPA simulations,
the ground-state SCF calculation for ($N-2$)-electron system is first performed,
then excitation energies are obtained as the differences between two-electron addition energies.
To reduce the computational cost,
the ppRPA equation is solved with the Davidson algorithm in an active space consisting of canonical orbitals, followed by an extrapolation scheme to obtain the full-space results.
We demonstrated that ppRPA provides a balanced description of correlated excited states in all tested defect systems, including VC in diamond, 
VO in MgO, and C$_{\text{B}}$C$_{\text{N}}$ in h-BN.
The errors from ppRPA@B3LYP for predicting vertical excitation energies of the tested point defects are mostly smaller than $0.1$ \,{eV}.
In particular,
ppRPA achieves good accuracy for excitation energies of VC in diamond with various geometries,
which is challenging for single-reference methods. This good performance is a result of seamless Fock space embedding in ppRPA, which captures explicitly the correlated interactions of two particles or two holes in the medium of the $N$-electron system described with a density functional approximation\cite{zhangAccurateEfficientCalculation2016}.
%As demonstrated in this work and our previous work\cite{liAccurateExcitationEnergies2024}, ppRPA treats $(N\pm2)$-excited states with a seamless Fock space embedding approach of capturing explicitly the correlated interactions of two particles or two holes in the medium of the $N$-electron system described with a density functional approximation\cite{zhangAccurateEfficientCalculation2016}.
Furthermore, 
we developed the NTO approach in ppRPA,
which provides important physical insights into electronic transitions in defect systems and the multireference character of associated states.
We conclude that ppRPA shows promise as a low-cost yet accurate tool for investigating excited-state properties of point defect systems.

\begin{acknowledgement}
T.Z. and J.L. are supported by the National Science Foundation (Grant No.~CHE-2337991) and a start-up fund from Yale University. 
J.L. also acknowledges support from the Tony Massini Postdoctoral Fellowship in Data Science. 
Y.J. acknowledges support from the Midwest Integrated Center for Computational Materials (MICCoM) as part of the Computational Materials Science Program funded by the U.S. Department of Energy, Office of Science, Office of Basic Energy Sciences, under Contract No.~DE-AC02-06CH11357. J.Y. and W.Y. acknowledge the support from the National Science Foundation (Grant No.~CHE-2154831).
\end{acknowledgement}

\section*{Supporting Information}
Details about geometry optimizations,
tabulated excitation energies of defects obtained from ppRPA,
basis set convergence in ppRPA,
tabulated excitation energies of defects obtained from TDDFT.

\section*{Data Availability Statement}
The data that support the findings of this study are available in the main text and the supporting information.

\bibliography{ref,unpublished}

\end{document}